\documentclass[twocolumn,
showkeys,
preprintnumbers,amsmath,amssymb,page]{revtex4}
\usepackage{graphicx}
\usepackage{dcolumn}
\usepackage{bm}

\begin{document}

\preprint{\small
Proceedings of 10th Quantum Information Technology Symposium (QIT10),
Gakushuin University, Tokyo, May 24-25, 2004}

\title{ {\large L}{\small OCATIONAL}  
{\large Q}{\small UBIT}  {\large R}{\small EALIZATION}  {\small WITH}  
{\large O}{\small NE-DIMENSIONAL} {\large C}{\small ONTACT}
 {\large I}{\small NTERACTIONS}
 }
\author{ Taksu Cheon
\footnote[1]{ 
        email: taksu.cheon@kochi-tech.ac.jp  \\
        http://www.mech.kochi-tech.ac.jp/cheon/
                  }
}
\affiliation{
Laboratory of Physics, Kochi University of Technology,
Tosa Yamada, Kochi 782-8502, Japan
}
\date{March 3, 2004}

\begin{abstract}
We show that the $U(2)$ group structure of thin barriers
can be adopted for quantum information processing when used 
in combination with environmental potential whose bouncing 
modes are profile preserving.
Qubits are realized as wave functions localized in either side of the 
barrier which divides the one-dimensional system into two regions.
It is argued that this model is a theoretical prototype of a robust and scalable
quantum computing device. 
\end{abstract}

\maketitle

\section{Introduction}

Quantum computation has emerged as one of the prime
source of inspiration
for mathematical quantum mechanics \cite{NC00}.  
It is a information processing based on quantum state
belonging to a $U(2)$ group.
Typically, spin one-half is considered as a natural playground.  
But any two level system can be utilized.
In a separate development, $U(2)$ structure has been 
uncovered \cite{SE86, TF00, CF01,TF01, FC03} 
in one dimensional system with 
generalized point interaction \cite{AG88, CS99, AK00}. 
A natural question is whether
it can be used for quantum information processing.

The purpose of this article is to show that that is indeed possible 
with the help of background potential, which move the particle back and forth
while keeping the wave function profile.  We consider controlling that motion 
by manipulating the properties of the barrier which separate the two spacial regions
of the system.  This quantum barrier is nothing but the point interaction whose characteristics
is specified by $U(2)$.  When we identify the wave function localized at
each of the two separated regions as qubits $\left|0\right>$ and $\left|1\right>$,
this $U(2)$ actually corresponds to the relevant operation of quantum computation.

The resulting model system takes the appearance of the quantum version of
that ancient eastern calculational device of {\it abacus}.  
We argue that this model could be a prototype of a robust quantum qubit device which 
excels in stability, controllability and scalability. 

\section{The Model}

We consider a one dimensional system
of quantum particle moving on $x$ axis subjected to a harmonic oscillator
potential of frequency $\omega = 2\pi/T$ and the inverse square potential 
with the strength $g$;
\begin{eqnarray}
\label{e1}
V(x)={ {\omega^2}\over {2} } x^2 + g { 1 \over {x^2}} .
\end{eqnarray}
With $g=0$, the background potential is reduced to the elementary harmonic oscillator.
In this article, we shall mainly work with this simple limit.
Then, we further add the generalized point interaction placed 
at the origin whose characteristics are controllable.  
We define boundary vector at $x \to +0$ and $x \to -0$ as
\begin{eqnarray}
\label{e2}
\Psi =
\begin{pmatrix}
  \psi(0_+) \\
  \psi(0_-)   
\end{pmatrix},
\quad
\Psi ' =
\begin{pmatrix}
  \psi ' (0_+) \\
 -\psi ' (0_-)   
\end{pmatrix} .
\end{eqnarray}
The point interaction is described by an element $U$ 
of  unitary group $U(2)$, that specifies the value of the boundary
vector such that 
\begin{eqnarray}
\label{e3}
( U - I ) \Psi + i ( U + I ) \Psi ' = 0 .
\end{eqnarray}
In other word, all point interactions allowable in quantum mechanics form
a family described by the set of four parameters, whose manifold structure is
given by $U(2) \simeq S^1 \times S^3$ \cite{FT00,MT02}.

The elementary example of $\delta$-interaction is but a very special one
parameter family within this wider class of interactions.
In fact, generalized point interaction $U(2)$ comprises such 
exotic interactions that cause discontinuity in the wave function itself, and
also the ones that have constant transmission probability which is
independent of the particle energy.  
%
%
\begin{figure}
\includegraphics[width=6cm]{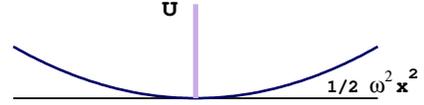}
\caption{\label{fig1}
The background harmonic potential and point interaction placed at the origin
as a barrier separating right and left regions.}
\end{figure}
%

Explicit construction of these highly singular point interactions
has been achieved in terms of singular short-range limits of 
known interactions \cite{CS98,EN01}.  
Here, we only illustrate some of the prominent examples.
The identity matrix $U = I$ results in $\psi '(0_+)=\psi '(0_-) = 0$, signifying the
inpenetrable barrier at $x=0$ with Neumann boundary at its both side.
Similarly, the negative ofidentity matrix $U=-I$ results 
in $\psi (0_+)=\psi (0_-) = 0$, another inpenetrable wall with Dirichlet boundary.
Curiously, it is  $U=\sigma_1$ that results in barrier-less free oscillater,
in $\psi (0_+)=\psi (0_-)$, $\psi '(0_+)=\psi '(0_-)$.  
An inportant example is the case of Hadamard matrix $U = H$ 
$= (\sigma_1+\sigma_3)/\sqrt{2}$.  It turns out that this point interaction lets the particle 
with all energy transmit by probability one-half.

Our central assertion is that the parametric $U(2)$ describing the whole
family of quantum point interaction can be turned into a {\it Hilbert space} $U(2)$
{\it of particle states} in the following manner:
If we identify the state localized at one side (let's say, right side)
as qubit $\left|0\right>$ and the other side as $\left|1\right>$, after half-period $T/2$, 
arbitrary state  $\left|\psi\right> = \alpha_0 \left|0\right>+ \alpha_1 \left|1\right>$ 
is turrned into $U\left|\psi\right>$.

Essential first step for the proof is the decomposition of
 $S^1 \times S^3$ into the spectral torus $T^2$ and the  isospectral
sphere $S^2$ \cite{TF01}.  The decomposition is realized, in terms of 
an element $U$ $\in U(2)$ as  
\begin{eqnarray}
\label{e4}
U=\sigma D \sigma
\end{eqnarray}
where
%
\begin{eqnarray}
\label{e5}
D=
\begin{pmatrix}
   e^{i \theta_+}  &   0 \\
    0  &   e^{i \theta_-}   
\end{pmatrix},
\\ \nonumber
\sigma=
\begin{pmatrix}
   \cos {\mu\over 2}  &  e^{i \nu} \sin {\mu\over 2}  \\
    e^{-i \nu}  \sin {\mu\over 2}  &   -\cos {\mu\over 2}   
\end{pmatrix} .
\end{eqnarray}
The range of the parameters are given by $0  \leqslant \theta_\pm <2\pi$,
$0 \leqslant \mu < \pi$ and $0 \leqslant \nu < 2\pi$. 
The parameter space $\{ \theta_+, \theta- \}$ forms a torus and
 $\{ \mu, \nu \}$  a sphere.  The latter is to be indentified with the 
Bloch sphere in quantum computation terminology.
A noteworthy feature of the system is that the energy levels are decided solely
$D$ and changing $\sigma$ will keep the spectra.

It is sufficient if we consider the operation of $D$ and $\sigma$ separately.  
We measure the phase of the states relative to the gound state of the
unperturbed harmonic oscillator $\exp{i \omega t}$. $U=D$ represents
the inpetrentrable hard wall,
 and $U=\sigma$ is the ``scale invariant" point interaction
which represents the energy (state) independent half-penetrating wall. 

\begin{figure}
\includegraphics[width=6.5cm]{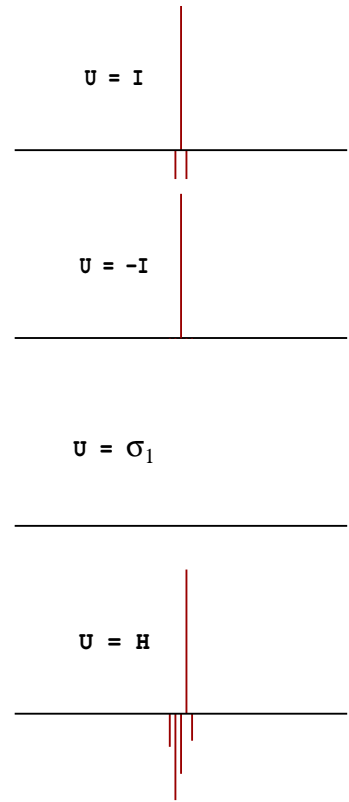}
\caption{\label{fig:epsart}
Illustrative representation of the construction of 
some of $U(2)$ barriers in terms of multiple $\delta$ functions placed in 
disappearing distances with diverging strength.
}
\end{figure}
%

%
\section{Scale Invariant Barrier}
First, we look at the the effect of the operation of $\sigma$ on wave functions.  
The connection conditions of wave function at the origin is given by
\begin{eqnarray}
\label{ei6}
\psi(0_+) = \lambda \psi(0_-), 
\quad 
\psi '(0_+) = {1\over \lambda^*} \psi '(0_-) ,
\\ \nonumber
{\rm where}\quad
\lambda=e^{i \nu}\sqrt{ {1+\cos{\mu\over 2}} \over  {1-\cos{\mu\over 2}} }.
\end{eqnarray}
The eigenvalues of the system are unchanged from the  free harmonic oscillator's
$\{n\omega\}$, and its eigenfunctions of the system $\{\chi^\lambda_n\}$  
are also analoguous to the free one $\{\chi_n\}$ having discontinuity at the origin 
and being expanded/shrank at one side;
\begin{eqnarray}
\label{ei7}
\chi^\lambda_{n}(x)=
N
\left[
\lambda \chi_{n}(|x|)\Theta(x)
+\chi_{n}(|x|)\Theta(-x)
\right]
\\ \nonumber
n = 0, 2, 4, \cdots ,
\\ \nonumber
\chi^\lambda_{n}(x)=
N
\left[
 \chi_{n}(|x|)\Theta(x)
- \lambda^* \chi_{n}(|x|)\Theta(-x)
\right] 
\\ \nonumber 
n = 1, 3, 5, \cdots ,
\end{eqnarray}
where the normalization is given by $N =$ $ \sqrt{ 2 / (|\lambda|^2+1) }$.
Arbitrary state $\psi(x,t)$ is represented as
\begin{eqnarray}
\label{ei8}
\psi(x,t) 
&=&
\sum_n A_n 
\chi^\lambda_n(x) 
e^{i n\omega t}
\\ \nonumber
&=&
{ {\lambda} \over {|\lambda|^2+1} }
\left[
  \lambda^* S(x) -  M(x) e^{i \omega t}
\right] \Theta(x) e^{i 2m \omega t}
\\ \nonumber
&+&
{ {\lambda^*} \over {|\lambda|^2+1} }
\left[ 
  S(x) +  \lambda M(x)e^{i \omega t}
\right] \Theta(-x) e^{i 2m \omega t} 
\end{eqnarray}
where we define
\begin{eqnarray}
\label{ei9}
S(x) &=& 
{2 \over {\lambda^* N} }
\sum_mA_{2m} \chi_{2m}(|x|) ,
\\ \nonumber
M(x) &=& 
-{ 2 \over {\lambda N} }
\sum_mA_{2m+1}\chi_{2m+1}(|x|) .
\end{eqnarray}
Let's assume that at $t=0$, the wave function is localized in the region $x>0$.
That is possible only if we have $M(x) =  -S(x) / \lambda$. 
We then have
\begin{eqnarray}
\label{ei10}
\psi(x,0)  \!\!\!
&=&  \!\!
\Theta(x) \  S(x)
\\ \nonumber
\psi(x,{T \over 2})  \!\!\!
&=&  \!\!
\left[ 
  \cos {\mu\over 2} \Theta(x) + e^{i \nu}\sin {\mu\over 2}  \Theta(-x) 
\right]  S(x) .
\end{eqnarray}
Similarly, with $S(x) =  M(x) / \lambda^*$, we have
\begin{eqnarray}
\label{ei11}
\psi(x,0) \!\!\!
&=& \!\!
\Theta(-x) \  M(x)
\\ \nonumber
\psi(x,{T \over 2}) \!\!
&=& \!\!
\left[ 
  e^{-i \nu}\sin {\mu\over 2} \Theta(x) - \cos {\mu\over 2}  \Theta(-x) 
\right]  M(x)  .
\end{eqnarray}
Thus, our assertion is proven for $\sigma$.
We note that the parameter values ($\mu = 0, \nu=0$), ($\mu = \pi, \nu=0$) 
and ($\mu=\pi/2 , \nu=0$) respectively correspond to Identity, Not and Hadamard
operations in quantum computation languages.
%
\begin{figure}
\includegraphics[width=6.5cm]{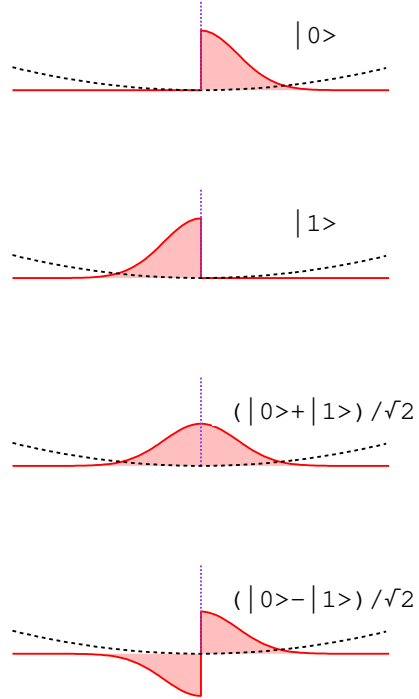}
\caption{\label{fig3}
The localized states to the right and the left of the barrier at the origin are
identified as the qubit state $\left| 0 \right>$ and $\left| 1 \right>$ respectively.
$U=I$ represents the closed gate, while $U=\sigma_1$ is the open gate case
in which they are interchanged in the period $T/2$. With Hadamard operator 
$U = H$, the mixed states  
$ (\left| 0 \right>\pm\left| 1 \right>)/\sqrt{2}$ are obtained 
from $\left| 0 \right>$ and $\left| 1 \right>$.
In all cases, the profile of the wave function is kept.
}
\end{figure}

\section{Hard Barrier}

Now, we look at $D$ which is the point interaction belonging to the torus
$\{ \theta_+, \theta_- \}$.  
The effect of $D$ on a qubit is applying a conditional phase
whose value depends on whether the qubit is $\left|0\right>$ or $\left|1\right>$.
\begin{eqnarray}
\label{eh12}
\sin {\theta_+ \over 2} \psi(0_+, t) + \cos {\theta_+  \over 2} \psi ' (0_+, t) = 0
\\ \nonumber
\sin {\theta_-  \over 2} \psi(0_-, t) - \cos {\theta_-  \over 2} \psi ' (0_-, t) = 0 .
\end{eqnarray}
This is nothing but two non-communicating sub-systems
separated by impenetrable barrier. 
It is sufficient to consider only one side, say, $x>0$ side, 
since the structure of the problem is the same. We take out the index $+$.
\begin{eqnarray}
\label{eh13}
\sin {\theta \over 2} \psi(0, t) + \cos {\theta \over 2} \psi ' (0, t) = 0 .
\end{eqnarray}
For $\theta = 0$, one has Neumann boundary, and
$\psi(x,t) = \psi(x,0)$
so the wave function stays the same 
(modulo ground state oscillation $\exp(i \omega /2 t)$).
For $\theta = \pi$, one has Dirichlet boundary, and
$\psi(x,t) = \psi(x,0) \exp(i \omega t)$.
so the wave function obtain phase $\pi$ after half period 
$T/2 = \pi / \omega$.
For generic $\theta$ between  $0$ and $\pi$, one has some 
``in between'' boundary condition which results in some mixture of 
$\psi_(x)$ with energy $\varepsilon(\theta)$ which should be 
\begin{eqnarray}
\label{eh14}
 0 < \varepsilon < \omega
\end{eqnarray}
So, after the half period,  the wave function obtain phase
\begin{eqnarray}
\label{eh15}
\psi(x,{T\over 2}) = \exp{ ( i \eta(\theta) \pi )} \psi(x,0),
\end{eqnarray}
where we define $\eta(\theta) = \varepsilon(\theta) / \omega$,
{\it provided} that the energy spectra of the system is {\it  equally spaced.}
The phase $\eta$ is a quantity satisfying $0< \eta < 1$, and
$\eta(\theta)$ should be some monotonous increasing function.  
Therefore, having $D$ applied for half-period, 
desired phase would be added.
The problem is, that 
the system have only {\it approximately}  equally spaced energy 
levels except for the case of large negative 
$g$ which is the strong attractive limit for the inverse square part of the
background potential.
So the strict phase operation is not  always possible 
with the application of corresponding $U(2)$ point interaction, 
except for the special cases.
The result of $D$ operation, in general, is therefore corrupted by 
the mixing of undesired higher harmonics with random phases.
However, there are simple workaround of this problem.
Instead of changing the wall property, we can c apply additional constant
potential $V_{ad}(x) = \eta \omega$ to obtain
the desired phase after half period $T/2$.
Another possibility is changing the harmonic oscillator frequency itself.
The former method seems to be simpler and practical.

As in any qubit realization, the choice of the basis is arbitrary, and its any unitary 
transformation is a legitimate basis. In quantum cryptography, for example, 
two bases connected by the Hadamard transformation is utilized.  Typically,
spin up/down basis and right/left basis are used.
In our case, that corresponds to the left/right localized basis 
and symmetric/antisymmetric basis.

\section{Prospectus}

In mathematical term, the two-qubit operation comprises $U(4)$ group.  
This is a natural extension to the $U(2)$, whose quantum wire realization 
has been the subject of this work.  Analogous realization of this $U(4)$ exists 
in the form of "quantum $X$-junction", a graph of four half lines whose edges 
are connected at single point \cite{EX96}.  
Thus, the study of quantum graphs 
now carries even more urgency.

A practical question in the experimental realization of our scheme is how to change 
the characteristics of the thin barrier representing the generalized point interaction. 
Preferably, it is to be a quantally operating device with triggering mechanism 
utilizing particles of far smaller mass or energy scale compared to the particle 
used as the qubit carrier. Once such device is constructed 
(no doubt that will be done, in time), and if that trigger is coupled to 
the presence absence of the qubit carrier in neighboring device, 
we will have a realization of two-qubit operations such as control-not in place.  
This type of setup would be quite a bit more advantageous in term of 
scalability compared to the qubit manipulation utilizing the forced transition 
by external laser beam.
%
%
%

One of the advantage of our implementation is the robustness of the
qubit due to the simplicity of the setup, which is matched
 only by the spin implementations.
In contrast, most solid-state based approaches use particle states which
could easily be lost in temperature fluctuations, whose suppression can
 be costly and potentially inhibiting in the setup requiring large number of qubits.
%
%
We hope that the model considered here could offer a basis for an alternative
location based quantum device which is simple, robust and truly scalable.

The content of the current work will be reported elsewhere in full \cite{CF04}.

We thank Dr. M. Iwata, Dr. Y. Kikuchi for helpful discussions. 

%


\end{document}